\documentstyle[amssymb,aps,epsfig]{revtex}
\newcommand {\pBe}      {\mbox{p+Be}}
\newcommand {\pCu}      {\mbox{p+Cu}}
\newcommand {\pAu}      {\mbox{p+Au}}

\begin{document}

\title{Antiproton Production in $p+A$ Collisions at AGS Energies}

%
%
\newcommand{\BNL}{$^{(1)}$}
\newcommand{\CU}{$^{(2)}$}
\newcommand{\FSU}{$^{(3)}$}
\newcommand{\IIT}{$^{(4)}$}
\newcommand{\IA}{$^{(5)}$}
\newcommand{\KSU}{$^{(6)}$}
\newcommand{\LBNL}{$^{(7)}$}
\newcommand{\LLNL}{$^{(8)}$}
\newcommand{\ORNL}{$^{(9)}$}
\newcommand{\SUNYS}{$^{(10)}$}
\newcommand{\UT}{$^{(11)}$}
\newcommand{\YU}{$^{(12)}$}

\author{
	I.~Chemakin,\CU\
	V.~Cianciolo,\LLNL\ORNL\
	B.A.~Cole,\CU\
	R.~Fernow,\BNL\
	A.~Frawley,\FSU\
	M.~Gilkes,\SUNYS\
	S.~Gushue,\BNL\
	E.~Hartouni,\LLNL\
	H.~Hiejima,\CU\
	M.~Justice,\KSU\
	J.H.~Kang,\YU\
	D.~Keane,\KSU\
	H.~Kirk,\BNL\
	M.~Kreisler,\LLNL\
	N.~Maeda,\FSU\
	R.L.~McGrath,\SUNYS\
	S.~Mioduszewski,\UT\BNL\
	D.~Morrison,\UT\
	M.~Moulson,\CU\
	N.~Namboodiri,\LLNL\
	G.~Rai,\LBNL\
	K.~Read,\UT\
	L.~Remsberg,\BNL\
	M.~Rosati,\BNL\IA\
	Y.~Shin,\YU\
	R.~Soltz,\LLNL\
	S.~Sorensen,\UT\
	J.~Thomas,\LLNL\LBNL\BNL\
	Y.~Torun,\SUNYS\BNL\IIT\
	D.~Winter,\CU\
	X.~Yang,\CU\
	W.A.~Zajc,\CU\
	and Y.~Zhang,\CU\
	}

%
%
\bigskip
\address{
\BNL\ Brookhaven National Laboratory, Upton, New York 11973\\
\CU\ Columbia University, New York, NY 10027 and Nevis Laboratories, Irvington, NY 10533\\
\FSU\ Florida State University, Tallahasee, FL 32306 \\ 
\IIT\ Illinois Institute of Technology, Chicago, IL 60616 \\
\IA\ Iowa State University, Ames, IA 50010 \\
\KSU\ Kent State University, OH \\
\LBNL\ Ernest O. Lawrence Berkeley National Laboratory, Berkeley, CA \\
\LLNL\ Lawrence Livermore National Laboratory, Livermore, CA 94550 \\
\ORNL\ Oak Ridge National Laboratory, Oak Ridge, TN 37831 \\
\SUNYS\ State University of New York at Stony Brook, Stony Brook, NY \\
\UT\ University of Tennessee, Knoxville, TN \\ 
\YU\ Yonsei University, Korea \\
}

\maketitle

\begin{abstract}
Inclusive and semi-inclusive measurements are presented for antiproton 
($\bar{p}$) production in proton-nucleus collisions at the AGS.
The inclusive yields per event increase strongly with increasing beam
energy and decrease slightly with increasing target mass. 
The $\bar{p}$ yield in 17.5~GeV/c p+Au collisions decreases with grey track
multiplicity, $N_g$, for $N_g>0$, consistent with annihilation within the
target nucleus.  The relationship between $N_g$ and the number of scatterings
of the proton in the nucleus is used to estimate the $\bar{p}$ annihilation
cross section in the nuclear medium.
The resulting cross section is at least a factor of five smaller than
the free $\bar{p}-p$ annihilation cross section when assuming a small or
negligible formation time.  Only with a long formation time can 
the data be described with the free $\bar{p}-p$ annihilation cross section. 
\end{abstract}

\twocolumn

\newcommand{\markcut}[1]{{}}
Antiprotons are a topic of great interest 
in Relativistic Heavy Ion Physics~\cite{kb,sb,bb,ab,rs} because  
enhanced production may result from the formation of the Quark
Gluon Plasma (QGP)~\cite{hs} and because antiprotons may
provide an experimental measure of the baryon density of matter produced in 
$A+A$ collisions due to the large $\bar{p}-p$ annihilation cross 
section~\cite{gg}.  Comparisons
of antiproton production in E802~\cite{ce802} and E878~\cite{e878}
with cascade models show that, in fact, 
more antiprotons are produced
in these collisions than would be expected from ordinary hadronic
production and the effects of final-state
absorption.  However, the same models have shown that these larger $\bar{p}$
yields could result from increased production and/or decreased absorption.
One proposed mechanism for increased production of $\bar{p}$ is 
the multiple scattering of the incident baryons~\cite{multistep}.  
Several mechanisms for decreased absorption have been proposed
including a finite $\bar{p}$ formation time, ``shielding'' of the absorption 
process~\cite{ARC}~(ARC),
and a time delay due to the formation of a $p-\bar{p}$ 
quasi-bound state~\cite{ppbarstate,ppbarmol}~(RQMD).  
Proton-nucleus collisions
provide a valuable tool for disentangling these
competing effects and elucidating the dynamics
of antiproton production and absorption in a nuclear environment
because the density of absorbers (nucleons) in the nucleus is
well understood. In particular, at energies close to the antiproton
production threshold, it is expected that most antiprotons will be produced
from the first scattering of the proton in the nucleus. The remaining
thickness of the nucleus may then simply act as an absorber. 
Previous measurements of $\bar{p}$ production in
$p+A$ collisions have been hindered by
poor statistics and the fact that most $p+A$ measurements
are inclusive.  Antiproton yields presented here are shown first
inclusively, comparing different beam energies and targets, 
and then as a function of the number of scatterings of
the projectile proton in the nucleus.
We focus on our semi-inclusive measurement to address
questions about the first-collision model and in-medium modifications to
the annihilation cross section. 

E910 is a TPC based $p+A$ experiment 
with downstream tracking, Cerenkov (CKOV), and time-of-flight (TOF) detectors.
The E910 apparatus has been described in detail elsewhere~\cite{e1},
but here we will again give a brief description of 
the time-of-flight (TOF) detector, used to identify antiprotons, and the
trigger used for this data set.
The TOF is located approximately 8~m from the target with an active area
of approximately 5$\times$2~m$^2$.  It is made up of 32 scintillating slats
with readout at the top and bottom of each slat.
Protons can be separated from pions and kaons by more than 2~$\sigma$ 
up to a momentum of 3.0~GeV/c and by more than 1~$\sigma$ up to 3.5~GeV/c,
where the timing resolution for protons is 164~ps.
The data presented here were collected with a scintillating fiber trigger.
The scintillating fiber was placed approximately 2~cm downstream of the 
target.  A minimum bias interaction is defined as an event having two hits in 
each of the two layers of the scintillating fiber, and a central interaction 
is an event with a total of 20 or more hits in the fiber.
We include only central triggers for which there were greater than 2 ``grey 
tracks'' in the event, where a grey track is defined as a ``slow'' proton or 
deuteron.  A slow proton has momentum $0.25<p<1.2$~GeV/c and a slow deuteron
has momentum $0.5<p<2.4$~GeV/c.

\begin{figure}
\begin{center}
\epsfig{file=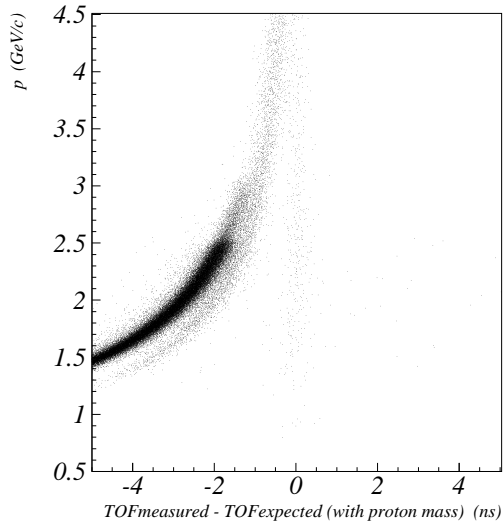,height=3.0in}
\end{center}
\caption{Total momenta for negative tracks vs. difference between measured flight time and expected flight time in ns.
\label{fig:pbarid}}
\end{figure}

For track quality, we require at least 10 hits in the 
time projection chamber (TPC) and that the track originated from the 
event vertex.  Although our primary particle identification for antiprotons
comes from time of flight, we additionally use Cerenkov information and 
ionization energy loss in the TPC to reduce background.  We
require that the ionization energy loss is within 3~$\sigma$ 
of the proton dE/dx.  In the relativistic rise region,
where the pion band separates from the proton,
the measured dE/dx must be greater than 1.5~$\sigma$ from
the pion dE/dx.  We also apply a cut on the Cerenkov ADC
which, on average, corresponds to requiring less than 0.35 photoelectrons.
The effect of these cuts on the background can be seen in 
Fig.~\ref{fig:pbarid}, which will be described later in the text.
Quality cuts on the hits on the time-of-flight wall (TOF) include a cut on
the difference in horizontal position between a projected track
and the center of a
hit TOF slat, and a cut on the energy deposited on the TOF slat.
We further require the projected track to have at least 5 hits in the 3 drift 
chambers located between the TPC and the TOF wall.
We determine the efficiencies of these cuts and correct for them 
in $y$ and $p_T$.
Tracks are matched to the TOF wall with an 90$\pm5$\% efficiency.
A single correction factor is applied uniformly over $y$ and $p_T$.
The 5\% systematic uncertainty in this correction is included in the
overall errors of the results.

Figure~\ref{fig:pbarid} shows the momentum 
dependence for negative tracks of the difference
between the measured time of flight and the expected time of
flight assuming the mass of a proton.  The cuts on this distribution 
are momentum dependent and range from 800~ps for
low momentum tracks to 200~ps for tracks with momenta between 3 and 3.5~GeV/c. 
A momentum-dependent background of the identified antiprotons is
calculated and subtracted, amounting to an overall
correction of approximately 5\%.
The data have been acceptance corrected within our $y-p_T$ coverage
and corrected for the efficiencies of our cuts as mentioned above.
Our coverage ranges from 10-800~MeV/c in $p_T$ and 1-2 in units of rapidity.
The acceptance is largest at low $p_T$ near rapidities between 1.6 and 2.0.
The acceptance is limited in the low $y$, high $p_T$ region by
the spatial coverage of the TOF wall, 
while the high $y$ region is limited by the upper momentum cut of 3.5~GeV/c.
Since a sample of identified protons, using loose TOF criteria, 
is relatively clean compared to a sample of loosely identified
antiprotons, the efficiency of the cuts
can be estimated by applying the cuts to the sample of identified protons. 
The overall calculated 
efficiency ranges from 60-80\%, decreasing with increasing $p_T$.
The cut on the difference in horizontal position of the projected 
track from the
TOF hit is most inefficient near the edges of the TOF wall, 
thus affecting primarily the low
$y$, high $p_T$ region.  The particle identification cuts (on dE/dx and
Cerenkov ADC), which reduce the background, 
affect the efficiency primarily for higher momentum 
antiprotons (high $y$, high $p_T$ region).
We neglect the inefficiency due to multiple hits in one TOF slat
because the slat occupancy is only approximately 3\%.
Even in central events, where the multiplicities are larger, the slat
occupancy is less than 3\% because the particles tend to shift back in
rapidity and thus out of the acceptance of the TOF wall.   
The data have also been corrected for trigger biases which are determined by 
examining beam-triggered events (unbiased events) 
in two dimensions, the total number of charged 
particles in the event and the number of grey tracks $N_g$ in the event.
Particularly in ``central'' interaction triggered events, 
but also in ``minimum bias'' interaction
triggered events, there is a bias against events with small numbers
of charged particles and small $N_g$.  This bias
can be determined from beam-triggered events, by comparing
the distribution of the number of charged particles and $N_g$ for
those beam-triggered events that also passed the conditions for an interaction
trigger to the unbiased distribution.
We estimate feeddown from antilambdas, by applying the $\bar{p}$
selection cuts to a set of antilambdas identified in the TPC, 
to be less than 5\%.
Final (raw) event statistics for each data set are shown in 
Table~\ref{tab:evstat}.  The data sets analyzed include 17.5~GeV/c momentum 
$p+Au$ collisions, 12.3~GeV/c $p+Au$,
12.3~GeV/c $p+Cu$, and 12.3~GeV/c $p+Be$.
The target thicknesses are 4.5\%, 3.1\%, and 2.0\% of the interaction length
for Be, Cu, and Au, respectively. 

Our measure of centrality is defined by
the number of projectile collisions $\nu$, which is
derived from the number of grey tracks $N_g$ in an event.
Slow protons and deuterons are identified by their measured
ionization energy loss in the TPC.  The
momentum cut on the protons is $0.25<p<1.2$~GeV/c and on the deuterons is
$0.5<p<2.4$~GeV/c.  For a class of events with a given number of grey tracks
$N_g$, we derive the mean number of collisions.  The details of
our method to determine $<\nu(N_g)>$ are described elsewhere~\cite{e1}. 

\begin{figure}
\begin{center}
\epsfig{file=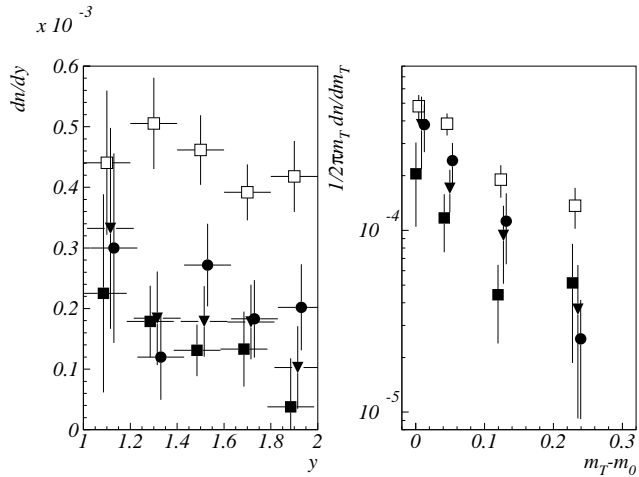,height=2.5in}
\end{center}
\caption{Target and beam momentum dependence of $p+A$ $\bar{p}$ spectra,
  a) $dn/dy$ distributions, b) $m_T$ distributions, $\bullet$ - \pBe,
  $\blacktriangledown$ - \pCu, $\blacksquare$ - \pAu\ at 12.3~GeV/c, 
  $\square$ - 17.5~GeV/c \pAu.  
  The points from different data sets are offset relative
  to each other in order to distinguish the error bars.
\label{fig:target_dep}}
\end{figure}

Antiproton yields for all 4 data sets are shown in Fig.~\ref{fig:target_dep}
as a function of rapidity and transverse mass.
The beam energy dependence is seen by comparing the two
$p+Au$ data sets.  The yields increase approximately by a factor of 3
from 12.3 to 17.5~GeV/c.  The yields for the different target sizes
can also be compared in this figure.
The transverse mass $m_T$ 
distribution for each target and beam momentum is fit to the following 
exponential,
\begin{equation}
{1 \over 2\pi m_T}{dn \over dm_T} = C_0 e^{{-(m_T-m_0)/T}},
\label{eq:ptfits}
\end{equation}
where $C_0$ and $T$ are fit parameters.
The results of the fits are tabulated in Table~\ref{tab:mtfits}.
Due to limited statistics, the errors on the fit parameters are large.  
However, comparing the inverse slope parameter for the 
largest data set, 17.5~GeV/c $p+Au$, with that of the 12.3~GeV/c $p+Be$
data set shows a significant increase from the smaller target to the larger
target.  Although the comparison is between two data sets with 
different beam energies,
such behavior is consistent with more reabsorption in a 
larger target.  The beam energy dependence of $<p_T>$, as 
previously measured in $p+p$ collisions~\cite{rossi}, cannot solely account
for the observed difference in slope.

\begin{figure}
\begin{center}
\epsfig{file=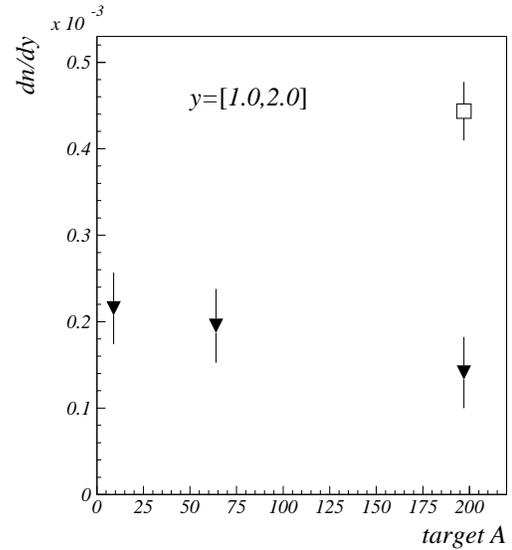,height=3.0in}
\end{center}
\caption{Target mass dependence of E910 $\bar{p}$ yields summed over $1<y<2$,
$\blacktriangledown$ -  yields for 12.3~GeV/c beam momentum,
$\square$ - 17.5~GeV/c.
\label{fig:int_yields}}
\end{figure}

The yields tend to increase with decreasing target size.
This trend is more evident when the yields are integrated
over the entire range of measured $y$ (1-2) and $p_T$ (10-800~MeV/c).
These yields, $dn/dy$ with y=1-2 and integrated over $p_T$, 
are shown in Fig.~\ref{fig:int_yields}.  There
is a $34 \pm 22$\% decrease in $dn/dy$ from the Be target to the Au target.  
The 17.5~GeV/c $p+Au$ yield is shown in the same figure for comparison.
The yield from $p+Au$ at 17.5~GeV/c is 3.1~times the yield at 12.3~GeV/c.
Although the likelihood of producing antiprotons may be greater in a
larger nucleus~\cite{kd}, the likelihood of reabsorption is also greater
due to the presence of more baryons.
At these beam energies, we find the effect of increased reabsorption
in the larger nucleus to be greater than any possible increase in production.

\begin{figure}
\begin{center}
\epsfig{file=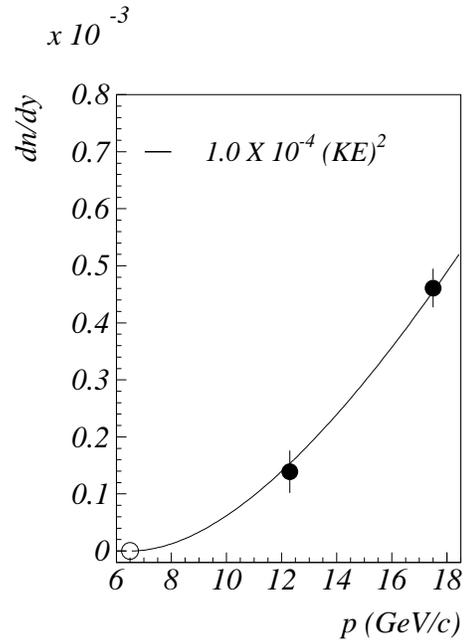,height=3.5in}
\end{center}
\caption{Energy dependence of E910 $p+Au$ $\bar{p}$ yields, $\bullet$ -
  $\bar{p}$ yields vs. beam momentum, $\circ$ - $\bar{p}$ production
  threshold.  The curve shows yields follow $KE^2$ dependence.}
\label{fig:ke2}
\end{figure}

The dependence of the yields on the beam momentum can be described by the 
available kinetic energy squared.  It was shown~\cite{pbar_kinedep} 
that antiproton yields for 
$p+p$ collisions at energies near the production threshold can be described
by 
\begin{equation}
(KE)^{2} = (\sqrt{s} - 4m)^{2},
\end{equation}
where $m$ is the mass of the antiproton.
In the reference, it is also shown (using phase-space arguments) 
that this dependence can be explained
by production through a three-body process rather than a four-body
process, indicating the possibility of an intermediate state.
Figure~\ref{fig:ke2} shows that this dependence also describes well our
$p+Au$ yields.  Having established this dependence, we can compare to data
at a different beam energy. 

\begin{figure}
\begin{center}
\epsfig{file=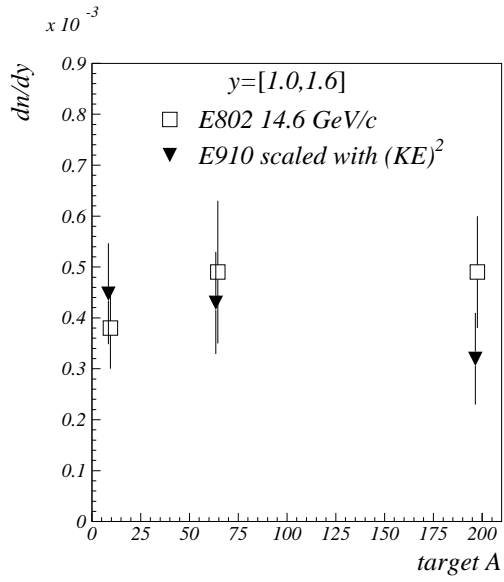,height=3.0in}
\end{center}
\caption{Comparison between E910 12.3~GeV/c $\bar{p}$ yields ($\blacktriangledown$) extrapolated to 14.6~GeV/c and E802 measurements ($\square$). See text for details.  The points from the 2 experiments are offset relative to 
each other in order to distinguish the error bars.
\label{fig:int_802yields}}
\end{figure}

Figure~\ref{fig:int_802yields} shows a comparison between our measurement of 
$dn/dy$ and the measurement by E802~\cite{e802}.  For the purpose of direct 
comparison, we have restricted our $y$ range from 1 to 1.6 and scaled the 
yields using $(KE)^2$ to correspond to the 14.6~GeV/c
beam momentum of E802.  Although the measurements are consistent with each 
other, E802 concluded no target dependence.  With increased statistics and a 
larger range in $y$ (Fig.~\ref{fig:int_yields}), we conclude that
there is indeed a modest target dependence at AGS energies.  

\begin{figure}
\begin{center}
\epsfig{file=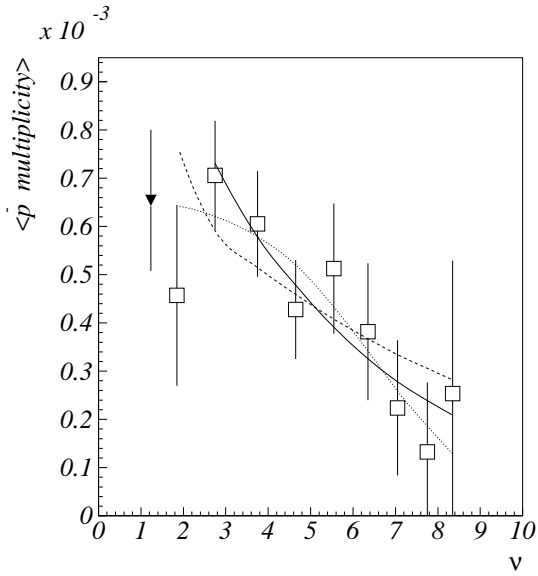,height=3.0in}
\end{center}
\caption{Dependence of 17.5 GeV/c \pAu\ $\bar{p}$ yields on
$\langle \nu (N_{\rm grey}) \rangle $ ($\square$).  Lines show results of absorption fits.  Fit~1 is shown by the dashed curve, fit~3 
the dotted curve, and fit~4 the solid curve  (see Table~\ref{tab:nufits} 
for details).  Also shown is 12.3~GeV/c \pBe\ data point extrapolated to 
17.5~GeV/c ($\blacktriangledown$).
\label{fig:au18_nu}}
\end{figure}

In addition to our inclusive measurements for different targets and
beam energies, we can use the dependence of the $\bar{p}$
multiplicity on centrality to help disentangle the mechanisms of 
production and reabsorption in the nucleus. 
The centrality (or $\nu$) dependence of $\bar{p}$ yields is shown in 
Fig.~\ref{fig:au18_nu}.  The antiproton yields are measured for each
value of $N_g$ and then plotted versus the mean $\nu$ for a given $N_g$.
Since the $N_g = 0$ bin corresponds to a mean of approximately 
two projectile collisions for $p+Au$, we have plotted the first
$N_g$ bin for $p+Be$ on the same figure (scaled from 12.3 to 17.5~GeV/c
beam momentum using the $(KE)^2$ relationship discussed above) simply as a 
reference of what the production may be in only one $p+N$ collision.
Although we may be somewhat biased against antiproton production
in the $N_g=0$ bin (the antiproton may be preferentially produced 
together with a slow proton due to baryon number conservation), 
it is possible that the increase from $N_g=0$ to $N_g=1$ 
is due to contributions to production beyond the first $p+N$ 
collision.  However, with the exception of the yield in the 
first $p+Au$ $N_g$ bin to the second, we see the mean antiproton multiplicity
decrease as the number of projectile collisions increases.
This relationship gives insight to
the amount of nuclear material traversed by the antiproton before it is 
reabsorbed.  With a few phenomenological assumptions, we present a 
quantitative measure of the survival probability of an antiproton as a 
function of the amount of nuclear material through which it passes.  
Because the beam energy is close to production threshold and the
antiprotons are strongly peaked at forward angles, we assume that
only the first collisions contribute to the production of
antiprotons, which are then assumed to follow the path of the
projectile through the nucleus.
Since we have conjectured possible contributions to production beyond the 
first collision, we will discuss the effect of such a modification to our
assumptions later in the text.
With this picture of $\bar{p}$ production, we quantify the 
reabsorption with the following equation,
\begin{equation}
\sigma(pA \rightarrow \bar{p}X) = \sigma(pp \rightarrow \bar{p}X) e^{-{\sigma_{abs} \over \sigma_{pN}}(\nu - 1)},
\end{equation}
where $\sigma(pp \rightarrow \bar{p}X)$ is the antiproton production
cross section for $p+p$, $\sigma_{abs}$ is the ``effective'' 
antiproton absorption cross section, and $\sigma_{pN}$ is the proton-nucleon 
interaction cross section.
Since the ``$\nu$'' plotted on the x-axis of
Fig.~\ref{fig:au18_nu} is simply an average value, $\bar{\nu}(N_g)$,
and each value of $N_g$ actually 
has a distribution of $\nu$ values associated with it, $P_{N_g}(\nu)$, we 
fold the above exponential with $P_{N_g}(\nu)$.
We determine $\sigma_{abs}$ by fitting
the $\nu$-dependent antiproton yields with the 
following function,
\begin{equation}
\sigma(pA \rightarrow \bar{p}X) = \sigma(pp \rightarrow \bar{p}X) P_{N_g}(\nu) e^{-{\sigma_{abs} \over \sigma_{pN}}(\nu - 1)}.
\label{eq:nu}
\end{equation} 
Folding the distribution, $P_{N_g}(\nu)$, for discretized 
values of $N_g$ results in a step-like behavior of the fit.  
We show smoothed fit functions in the figure.
The results of the fits with various sets of assumptions
are shown in Table~\ref{tab:nufits}.  We have done the fits with and without
allowing for a formation time,
\begin{equation}
\tau_{form} = {\nu_{form} \lambda \over \gamma v},
\end{equation}
during which the antiproton cannot annihilate.
In this case, the exponent in Eq.~\ref{eq:nu} becomes $-{\sigma_{abs} \over \sigma_{pN}}(\nu-\nu_{form}-1)$ for
$\nu > \nu_{form} + 1$, and there is no absorption ($\sigma_{abs}=0$) for
$\nu \le \nu_{form} + 1$.
One should note that the linear relationship between $\nu_{form}$ and 
$\tau_{form}$ is not as straightforward for very large values of $\nu_{form}$
which rely on fluctuations in the nuclear density distribution.
The quantities that are used to
calculate $\sigma_{abs}$ and $\tau_{form}$ from the fit parameters 
are a mean free path $\lambda$ of 2~fm, a proton-nucleon 
interaction cross section
$\sigma_{pN}$ of 30~mb, and a free annihilation cross section $\sigma_{ann}$ of
38~mb (at the mean measured momentum of 2.5~GeV/c 
for the antiprotons we detect).  Using a momentum of 2.5~GeV/c, we
calculate $\gamma$ and $v$.  
In the first 3 fits shown in the table, 
we include the first $p+Au$ data point ($N_g=0$) in the fit,
and in the next set of 3 fits, we
do not include this point (because of the initial increase
in yield from $N_g=0$ to $N_g=1$).  In addition to removing a
possible bias in the first data point from the fit, this also allows
for production beyond the first collision up to the value of $\nu = 2.4$ 
where the fit begins. 

The first row shown in Table~\ref{tab:nufits} is the result of a fit
assuming that the formation time is negligible.
The extracted $\sigma_{abs}$ is significantly reduced relative to 
$\sigma_{ann}$ (almost by a factor of 10).  This fit is shown in 
Fig.~\ref{fig:au18_nu} as a dashed curve.  In the following 2 fits, we
investigate the effect of a formation time $\tau_{form}$ on this result.
With no constraints on the fit, $\tau_{form}$ is very large.
However, with excessively large $\tau_{form}$ we lose the ability to uniquely
determine $\sigma_{abs}$ from the fit.
Thus, we constrain $\tau_{form}$ and $\sigma_{abs}$ separately in 2 fits.
Typical values used in transport models for the formation time
are 1-2~fm/c~\cite{bu}.  
Constraining $\tau_{form}$ to such values results
again in a reduced absorption cross section.  The fit parameters shown in fit~2
in the table are for a constraint of $\tau_{form} = 1$~fm/c, with 
which one obtains a $\sigma_{abs} = 4.6 \pm 0.9$~mb.  (This fit looks similar
to the fit with no formation time and is, therefore, not shown in the figure.)
The other possibility is to constrain the absorption cross section to be
equal to the free annihilation cross section.  Such a constraint leads to
$\nu_{form} = 6.7 \pm 0.7$, which corresponds to a
long formation time of $4.9 \pm 0.5$~fm/c or a formation length of 
approximately 13~fm in the nuclear rest frame.
Again, such large values of $\nu_{form}$ rely on density
fluctuations and thus do not have such a well-defined relationship to 
$\tau_{form}$.
The fit is shown by the dotted curve in Fig.~\ref{fig:au18_nu}.
The next 3 rows repeat the 3 fits described above, excluding the first
$p+Au$ data point from the fit.   
The results are qualitatively similar to those when including the first data
point.  Fit~4 in the table, which is shown with a solid curve in the figure, 
again assumes no formation time and results in a reduced
absorption cross section (approximately 5 times smaller than $\sigma_{ann}$). 
When including a formation time of 1~fm/c, fit~5 in the table,
$\sigma_{abs}$ is still reduced by approximately a factor of 5.
Finally, fit~6 shows that constraining $\sigma_{abs} = \sigma_{ann}$ results
in a long formation time, even when excluding the first data point.
The large discrepancy between $\sigma_{abs}$ and $\sigma_{ann}$, 
as derived from our simple model when the formation time is negligible or small
(1~fm/c), suggests that the ``effective'' 
annihilation cross section is very different from the free
annihilation cross section due to in-medium effects.  On the other hand,
the data can be described by the free $p-\bar{p}$ annihilation cross
section and a very long formation time.
A possible explanation for such a result is the formation of
an intermediate state where $\tau_{form}$ can be interpreted 
as the mean lifetime of the state.
With such a large $\tau_{form}$, the antiproton is born only in the
late stage of the propagation of this state through the nucleus, 
leaving little opportunity to get reabsorbed.  
This hypothesis could
be tested by measuring the absorption of other antibaryons which could
proceed through the same intermediate state.

In conclusion, we find the yields dramatically increase with increasing beam
energy and can be described by a dependence on the available kinetic energy 
squared.  The observed energy dependence can be understood if the
antiproton is produced through the decay of an intermediate 
state~\cite{pbar_kinedep}.  
We find a moderate decrease with increasing target mass, 
$34 \pm 22$\% from Be to Au
for beam momentum 12.3~GeV/c.  Finally, we have quantified the survival
probability of an antiproton in the nuclear medium as a function of the
number of collisions.  With this relationship and the assumption of a 
negligible or small formation time of 1~fm/c,
we find that the annihilation cross section is 
greatly modified within the nuclear medium, and that the ``effective''
absorption cross section is a small fraction of the free annihilation cross
section.  On the other hand, a full $p-\bar{p}$ annihilation cross section
would require a much longer formation time than normally assumed. 
Previous attempts to explain a suppression of the
annihilation of antiprotons within the nucleus include an increased 
hadron formation time~\cite{ce802,vais}, the formation of a $p-\bar{p}$
molecule~\cite{ppbarstate,ppbarmol} with a finite lifetime, 
and a ``shielding'' effect
due to the presence of mesons~\cite{ARC}.  All of such
phenomenological arguments manifest themselves as a delay
time during which the antibaryon cannot annihilate with a baryon.  
Shielding, in particular, is dependent on the density of the nuclear
medium and would probably not be a large effect in proton-nucleus collisions.
Our $\nu$-dependent yields, however, show that the reabsorption of
antiprotons is already greatly suppressed in $p+A$ collisions.
Production through an intermediate state, which does not get absorbed like
an antiproton, could also explain a
suppression of the annihilation of antiprotons.  
In conclusion, we observe a deviation from the expectations of the 
naive first-collision model in which the antiproton is produced on-shell 
in the first collision with a small or negligible formation time 
and then interacts
with nucleons according to the free $p-\bar{p}$ annihilation cross section.

We wish to thank R.~Hackenburg and the MPS staff, J.~Scaduto and
G.~Bunce for their support during E910 data-taking. We also thank Thomas Kirk,
BNL Associate Director for High Energy and Nuclear Physics, for his support of
our physics program.

This work has been supported by the U.S. Department of Energy 
under contracts with BNL (DE-AC02-98CH10886), Columbia (DE-FG02-86ER40281), 
ISU (DOE-FG02-92ER4069), KSU (DE-FG02-89ER40531), LBNL (DE-AC03-76F00098),
LLNL (W-7405-ENG-48), ORNL (DE-AC05-96OR22464) 
and UT (DE-FG02-96ER40982) and the National Science Foundation under 
contract with the Florida State University (PHY-9523974).

\bibliographystyle{prsty}

\begin{table}[h]
\begin{center}
\caption{Final (raw) event statistics.}
\vspace{0.5cm}
\begin{tabular}{|l|c||c|c|c|c|} \hline
Target & p & No. min. & No. central & No. $\bar{p}$ & No. $\bar{p}$ \\
 & (GeV/c) & bias triggers & triggers & min. bias & central \\
\hline
Au & 17.5 & 2.66~M & 1.06~M & 346 & 93 \\
\hline   
Au & 12.3 & 1.69~M & 0.46~M & 73 & 6 \\
\hline   
Cu & 12.3 & 1.26~M & 0 & 84 & 0 \\
\hline   
Be & 12.3 & 1.41~M & 0 & 102 & 0 \\
\hline   
\end{tabular}
\label{tab:evstat}
\end{center}
\end{table}

\begin{table}[h]
\begin{center}
\caption{Fit parameters of exponential fits to transverse mass distributions.}
\vspace{0.5cm}
\begin{tabular}{|l|c||c|c|} \hline
Target & p (GeV/c) & $C_0$~(GeV$^{-2}$c$^2)$ & T (MeV/c) \\
\hline
Au & 17.5 & $4.90~\pm~0.62 \times 10^{-4}$ & $157 \pm 34$ \\
\hline   
Au & 12.3 & $1.85~\pm~0.91 \times 10^{-4}$ & $98 \pm 58$ \\
\hline   
Cu & 12.3 & $2.84~\pm~0.95 \times 10^{-4}$ & $108 \pm 47$ \\
\hline   
Be & 12.3 & $4.10~\pm~0.92 \times 10^{-4}$ & $86 \pm 19$ \\
\hline   
\end{tabular}
\label{tab:mtfits}
\end{center}
\end{table}

\begin{table}[h]
\begin{center}
\caption{Fit parameters of antiproton absorption fits.  Fits~1-3 
include the first $p+Au$ data point in the fit, and fits~4-6 do not.}
\vspace{0.5cm}
\begin{tabular}{|l|c|c|c|c|} \hline
Fit & $\sigma_{abs} (mb)$ & $\tau_{form} (fm/c)$ & Constraint & $\chi^{2}/NDF$ \\
\hline
1 & 4.0 $\pm$ 1.6 & 0 & $\tau_{form}$=0 & 6.994/7 \\
2 & 4.2 $\pm$ 1.6 & 1 & $\tau_{form}$=1 & 6.562/7 \\
3 & 38 & 4.9 $\pm$ 0.5 & $\sigma_{abs}$=38 & 3.149/7 \\
\hline
4 & 6.9 $\pm$ 2.2 & 0 & $\tau_{form}$=0 & 2.005/6 \\
5 & 6.9 $\pm$ 2.4 & 1 & $\tau_{form}$=1 & 1.995/6 \\
6 & 38 & 4.7 $\pm$ 0.7 & $\sigma_{abs}$=38 & 1.928/6 \\
\hline   
\end{tabular}
\label{tab:nufits}
\end{center}
\end{table}

\clearpage

\clearpage 

\end{document}